\documentclass[aps,twocolumn,superscriptaddress,nofootinbib,tighten]{revtex4}
\def\x{{\bf x}}
\usepackage[bookmarks=false]{hyperref}
\usepackage{graphicx}
\begin{document}

\title{The likelihood for supernova neutrino analyses}
\author{A. Ianni} 
\affiliation{INFN, Laboratori Nazionali del Gran Sasso, Assergi (AQ),
  Italy}

\author{G. Pagliaroli}
\affiliation{INFN, Laboratori Nazionali del Gran Sasso, Assergi (AQ),
  Italy}
\affiliation{Dipartimento di Fisica, University dell'Aquila,  Coppito (AQ), Italy}

\author{A. Strumia}
\affiliation{Dipartimento di Fisica dell'Universit\`a di Pisa and
  INFN, Italy}
\affiliation{CERN, Theory division, CH-1211, Geneva 23, Switzerland}

\author{F. R. Torres}
\affiliation{INFN, Laboratori Nazionali del Gran Sasso, Assergi (AQ),
  Italy}
\affiliation{Instituto de F\'\i{}sica ``Gleb Wataghin'', 
Universidade Estadual de Campinas, UNICAMP,
Campinas, SP, Brasil}

\author{F. L. Villante}
\affiliation{Dipartimento di Fisica, University dell'Aquila,  Coppito (AQ), Italy}
\affiliation{INFN, Laboratori Nazionali del Gran Sasso, Assergi (AQ),
  Italy}

\author{F. Vissani}
\affiliation{INFN, Laboratori Nazionali del Gran Sasso, Assergi (AQ), Italy}

\begin{abstract}
We derive the event-by-event
likelihood that allows to extract the complete information
contained in the energy, time and direction of supernova neutrinos,
and specify it in the case of SN1987A data. We resolve discrepancies in
the previous literature, numerically relevant already in the 
concrete case of SN1987A data.

\centerline{\tiny Preprint: LNGS/TH-01/09 and CERN-PH-TH/2009-115}
\end{abstract}
\pacs{}

\maketitle

\section{Introduction}
SN1987A neutrino events \cite{exp} 
prompted many dedicated analyses.
Even if the number of detected neutrinos is limited,
these analyses provide interesting limits on neutrino  properties
and clues on the core collapse mechanism.

The question of which likelihood
should be adopted for supernova neutrino data analysis
will become crucial after the next galactic supernova,
when a much larger number of neutrino events will be collected.
These events will carry information on neutrino properties 
mixed with information about supernova properties,
so that we will need to jointly study their energy, time and 
direction distributions to try to extract all the relevant 
pieces of information. Therefore it will remain unpractical 
to bin the events and an event-by-event likelihood will remain 
the best tool for data analysis.

We present here the likelihood that should be ideally adopted for supernova 
neutrino data analysis.
Our likelihood is more general than those already present in the
literature~\cite{bahcall,ll89,neubig}-\cite{ll}. 
Moreover, we resolve discrepancies in the previous literature, 
numerically relevant already in the concrete case of SN1987A data.
We argue, in particular,  that the analysis of SN1987A 
neutrino data by Lamb and Loredo~\cite{ll} (LL), quoted since 2004
in the summary table of the Particle Data Group~\cite{pdg},
uses a likelihood that  incorrectly biases 
the analysis in favor of low energy events.
We here present the correct likelihood, 
generalizing the `traditional' form, 
advocated, e.g., by Jegerlehner,
Neubig and Raffelt~\cite{neubig}.

%\medskip

The structure of this paper is the following.
In Sect.~\ref{prim} we derive the
general form of the likelihood.
The application to a specific case of interest is discussed
in Sect.~\ref{thir}. Finally, in Sect.~\ref{four}
we compare our likelihood with other forms
adopted for the analysis of SN1987A neutrinos,
showing how the fitted parameters got biased.

\section{Derivation of the likelihood\label{prim}}
\subsection{General form of the likelihood}
We write the expected event number in the $i$-th bin as:
\begin{equation}
n_i= dt_i ~d\x_i\ \frac{dN}{dt \,d\x}(t_i,\x_i), \label{eq1}
\end{equation}
where $t_i$ represents the time coordinate, while $\x_i$ indicates the set of all other
observables (energy, position, direction, etc.) which define the properties of the $i$-bin.
We suppose that the bin sizes $dt_i~ d\x_i$ are infinitesimally small so 
that the condition $n_i\ll 1$ holds true:
therefore the probability that multiple events are collected in one bin is negligible and, thus, 
observing $N_{\rm ev}$ events corresponds to $N_{\rm ev}$ bins with 1 event,
and all other bins with 0 events.

According to Poissonian statistics (see e.g.,\ Appendix A of \cite{jhep})
the associated  likelihood is:
\begin{equation}
\mathcal{L }=
 \exp\left[ - \sum_{j=1}^{N_{\rm bin}}  n_j  \right]  \times \prod_{i=1}^{N_{\rm ev}}
 n_i,
 \label{l1}
 \end{equation}
where the sum in the exponent runs over all $N_{\rm bin}$ bins and
gives the total number of expected events, while the product runs
over all $N_{\rm ev}$ observed events. As usual, one can convert
this into a $\chi^2$ distribution as $\mathcal{L} = e^{-\chi^2/2}$.

\subsection{Distinguishing between signal and background}
Let us consider the case when the detected events are due to a
signal $S$, reprocessed in the detector through a response
function ${\cal R}$, and to a known (measured) background process $B$.
We have:
\begin{equation}
 \frac{dN}{dt d\x}(t,\x)\! \! =B(t,\x)\! +\! \int\! dt'd\x' S(t',\x')
 \mathcal{R}(t',\x',t,\x)
 \label{dng}
\end{equation}
The second term in the r.h.s. takes into account that a signal
produced at the time $t'$ and with coordinates $\x'$, due to
detector response, could be observed with a probability
$\mathcal{R}(t',\x',t,\x)$ at a different time $t$ and coordinate~$\x$.

By integrating over all possible detection times and coordinates,
we introduce the general form of the detection efficiency:
\begin{equation}
\eta(t',\x')\equiv \int dt~d\x~\mathcal{R}(t',\x',t,\x).
\label{edef}
\end{equation}
The efficiency obeys the condition $0\le \eta \le 1$,
if we describe a situation when the events can be lost.
By factoring out $\eta$ we define the
 smearing (or error) function~$\mathcal{G}$
\begin{equation}
\mathcal{G}(t',\x',t,\x) \equiv
\mathcal{R}(t',\x',t,\x)/\eta(t',\x') \label{gdef}
\end{equation}
normalized to unity:
\begin{equation}
\int dt~d\x~ \mathcal{G}(t',\x',t,\x)=1.
\end{equation}
The  background $B$, the efficiency $\eta$ and the smearing
$\mathcal{G}$ describe the experimental apparatus. Assuming that
they are known, we can use an experimental result to learn on the
signal $S$,  by the study of the likelihood function of
Eq.~(\ref{l1}) together with~(\ref{dng}).

\subsection{Simplifications}
In the case of interest, it is possible to further
simplify the problem by relying on the following assumptions:

%\bigskip

$(i)$ We assume that the response function
factorizes in the time and in the coordinates as follows
\begin{equation}
\mathcal{R}(t',\x',t,\x) =  r(t',t) \mathcal{R}(\x',\x).
\end{equation}
We introduce the time-independent
efficiency in the observables $\eta(\x')$, defined
in analogy to Eq.~(\ref{edef}):
\begin{equation}
\eta(\x')\equiv \int d\x~ \mathcal{R}(\x',\x) , \label{www}
\end{equation}
and the smearing function defined in analogy with
Eq.~(\ref{gdef}):
\begin{equation}
\mathcal{G}(\x',\x) \equiv \mathcal{R}(\x',\x)/\eta(\x').
\label{def11}
\end{equation}
Again, it is normalized to unity:
\begin{equation}
\int d\x ~\mathcal{G}(\x',\x)=1. \label{petain} \label{norm}
\end{equation}
We will discuss later the specific form of these expressions for SN1987A.

%\bigskip

$(ii)$ If the time $t$ is measured with negligible error, we have
\begin{equation}
r(t',t) =  \delta(t-t'), \label{errrr}
\end{equation}
possibly multiplied by a window function $w(t)$ to account 
for the dead time $\tau$ after an event, 
due to supernova or to background (for example, a muon), 
has been recorded. Concerning SN1987A data, only the
{\em relative} time between events of the detectors was measured
precisely; one needs to take into account the uncertainty in the
absolute time of the Kamiokande-II and Baksan events.

%\bigskip

$(iii)$ We can finally assume that the background does not depend on the time, namely
 \begin{equation}
B(t,\x) =  B(\x)
\label{btdef}
\end{equation}
possibly, multiplied by $w(t)$ to take into account for the
absence of any events, including those due to background,
during dead time. Eq.~(\ref{btdef}) implies that the background
can be {\em measured} in the period
when the signal is absent (as
for SN1987A).

%\bigskip

With these assumptions, Eq.~(\ref{dng}) simplifies to:
\begin{equation}
 \frac{dN}{dt d\x}(t,\x) =B(\x) + \int d\x'
\mathcal{G}(\x',\x) \eta(\x') S(t,\x').
 \label{dngp}
\end{equation}
Then, assuming that the $N_{\rm ev}$ events $\x_i$ have been measured at time $t_i$,
the likelihood in  Eq.~(\ref{l1}) becomes:
\begin{equation}
\begin{array}{l}
\mathcal{L}=
 e^{- \int dt d\x B(\x) - \int dt d\x' \eta(\x') S(t,\x')   } \times \\
\ \ \ \ \ \ \prod_{i=1}^{N_{\rm ev}}
 \left[ B(\x_i) + \int d\x' ~\mathcal{G}(\x',\x_i) \eta(\x')
S(t_i,\x') \right] dt_i d\x_i,
\end{array}
 \label{l2}
\end{equation}
where, in the exponent, we replaced the sum over all infinitesimal
bins with an integral and used~(\ref{petain}). By dropping 
constant factors, that are irrelevant for estimating the
parameters that control the theoretical expression of the signal
rate $S$, and replacing $\x'$ with $\x$, we~get
\begin{equation}
\begin{array}{l}
\mathcal{L}=
 e^{- \int dt~d\x ~ \eta(\x) S(t,\x)   } \times \\
\ \ \ \ \ \
\prod_{i=1}^{N_{\rm ev}}
 \left[ B(\x_i) + \int d\x ~\mathcal{G}(\x,\x_i) \eta(\x) S(t_i,\x)
 \right].
\end{array}
 \label{l3}
 \label{alfin}
\end{equation}
This form of the likelihood is general enough
for the purpose of analyzing SN1987A neutrinos.
Moreover, this is a generalization of the likelihood 
advocated in \cite{clive} for the study of radioactive decays, 
when the time of occurence of each event is measured.

As we already discussed, the 
dead time can be taken into account by extending the
time integral in the exponent only to
the time when the detector is on, thereby removing
the time intervals where data taking was 
stopped after each candidate signal event.
As long as $\tau$ is small enough, one can equivalently take into 
account the dead time due to background events by multiplying the 
integrand in the exponent of~(\ref{alfin}) by the 
average live-time fraction, $1 - \tau b_\mu$, where $b_\mu$ 
is the time-averaged background event rate.
Compare it with the discussion of \cite{kraus}, 
further elaborated in~\cite{ll}.

\section{Application to Kamiokande-II\label{thir}}
In order to specify the general formul\ae,
we choose a concrete and important example: we discuss
the likelihood for the water \v{C}erenkov detector Kamiokande-II.

\subsection{Generalities}
In this subsection, we collect some useful definitions.

The variables that characterize an event are:
\begin{equation}
\x_i=\{
E_i\mbox{ (energy) , }
\hat{n}_i\mbox{ (direction) , }
\vec{r}_i\mbox{ (position) } \}.
\label{esamp}
\end{equation}
For concreteness, we consider events resulting from the reaction
$\bar\nu_e p\to n e^+$ when a positron is detected through its
\v{C}erenkov light; similar considerations apply to the elastic
scattering reaction or the charged current reactions with nuclei.

In the construction of the likelihood 3 different directions 
are relevant:
the direction $\hat{n}_*$ of SN1987A; 
the {\em reconstructed} direction $\hat{n}_i$ of
each event; the {\em true} direction $\hat{n}$ of the positrons produced 
by the  detection process. For each event, the first 2 directions are 
fixed, while we have to integrate on the true direction of 
the positron, taking into account the detector response and the reconstructed 
event direction, as described in Eq.~(\ref{alfin}). 
To do this, it is convenient to use an ``event-centric'' system in which:\\
1) The reconstructed positron direction is along the $z$ axis,
\begin{equation}
\hat{n}_i=(0,0,1).
\end{equation}
2) The true positron direction is in the generic
direction:
\begin{equation}
\hat{n}=(\sin\theta \cos\varphi,\sin\theta
\sin\varphi,\cos\theta), \label{est}
\end{equation}
so that
\begin{equation}
\cos\theta=\hat{n}_i \hat{n}.
\label{eid}
\end{equation}
Thus, $\theta$
is the opening angle around
the reconstructed direction and $\varphi$ is the azimuthal angle.
The experimental collaborations usually
quote the error on the angle $\delta \theta_i$ between the true and the reconstructed direction
for each bin (rather than the the error on the direction versor $\delta \hat{n}_i$ itself).
\\
3) Finally, we have the versor $\hat{n}_*$ pointing
in the direction of the supernova.
This, without loosing in generality,
can be chosen in the plane $x-z$:
\begin{equation}
\hat{n}_*=(\sin\theta_i,0,\cos\theta_i), \label{nst}
\end{equation}
so that
\begin{equation}
\cos\theta_i=\hat{n}_*\hat{n}_i.
\end{equation}

\subsection{Smearing function}

In the simplest approximation, we can describe the 
smearing function by assuming that it factorizes 
according to: 
\begin{equation}
\mathcal{G}({\bf x},{\bf x}_i)= G_1(E\!-\!E_i,\sigma_1)
G_2(\hat{n}\!-\!\hat{n}_i,\sigma_2)
G_3(\vec{r}\!-\!\vec{r}_i,\sigma_3)
\end{equation}
where we denote by
\begin{equation}
G_n(\vec{x},\sigma)=\frac{\exp(-\vec{x}^{2}/2\sigma^2)}{N_n
(\sqrt{2\pi}\sigma)^n}, \label{cesent}
\end{equation}
a standard Gaussian in $n$ dimensions. We include a normalization factor 
$N_n$ to describe the presence of physical boundaries, like e.g., 
the fact that $\theta\in[0,\pi]$ and $\varphi\in[0,2\pi]$ when we integrate 
over the possible directions of $\hat{n}$.

The quantities $\sigma_{1,2,3}$ are functions 
of the variables of Eq.~(\ref{esamp}). We
identify $\sigma_{1,2,3}$ in the point $E=E_i,\hat{n}=\hat{n}_i,\vec{r}=\vec{r}_i$ 
with the error for the $i$-th event 
quoted by the experimental collaborations, e.g., 
\begin{eqnarray}
\nonumber
\sigma_1(E_i,\hat{n}_i,\vec{r}_i) &=& \delta E_i\\
\nonumber
\sigma_2(E_i,\hat{n}_i,\vec{r}_i) &=& \delta n_i\\
\sigma_3(E_i,\hat{n}_i,\vec{r}_i) &=& \delta r_i
\label{errors}
\end{eqnarray}
thus, we further approximate the smearing function:
\begin{equation}
\mathcal{G}({\bf x},{\bf x}_i)\approx G_1(E\!-\!E_i,\delta E_i)
G_2(\hat{n}\!-\!\hat{n}_i,\delta n_i)
G_3(\vec{r}\!-\!\vec{r}_i,\delta r_i)
\end{equation}
in the vicinity of each point where the likelihood should be 
evaluated --- see Eq.~(\ref{alfin}).\footnote{
An (arguably) more refined approximation that takes 
into account the Poisson nature 
of the photoelectron detection can be obtained 
by multiplying the constant errors 
$\delta E_i$, $\delta n_i$ and $\delta r_i$, 
by $\sqrt{E_i/E}$. We note, in this respect, 
that the normalization condition in Eq.~(\ref{norm}) 
is obeyed even in the general case, 
when the functions $\sigma_{1,2,3}$ vary with the 
true coordinates of positron, since
one integrates over the {\em reconstructed} 
coordinates.}

\subsection{Remarks on the angular distribution}

The expressions for energy and position smearing functions 
are essentially standard and do not need particular attention. 
The angular distribution requires, instead, a more detailed discussion.

First, we discuss the connection between 
the error $\delta n_i$ to be inserted into Eq.~(\ref{errors}) and 
the 1-sigma error on angle $\delta\theta_i$ indicated by 
the experimental collaborations. 
In the Gaussian assumption, the function $G_2$ 
can be written as:
\begin{equation}
G_2(\hat{n}\!-\!\hat{n}_i,\delta n_i) d\hat{n}=
\frac{d\varphi}{2 \pi}\times dc \frac{d\rho}{dc}(c,\delta n_i) 
\end{equation}
where the function $d\rho/dc$, given by: 
\begin{equation}
\frac{d\rho}{dc}=\frac{1}{N_2 \; \delta n^2_i}\exp\! \left[ \!-\frac{1-c}{\delta n^2_i}\right],
\label{aaa1}
\end{equation}
describes the distribution of the angle $c=\cos\theta=\hat{n}_i \hat{n}$ between the 
true and the reconstructed direction; the azimuthal angle $\varphi$ 
is uniformly distributed as appropriate for an unbiased detector. 
The normalization factor $N_2$ can be explicitely calculated
\begin{equation}
N_2 = 1-\exp(-2/\delta n^2_i),
\end{equation}
and it is very close to one for the typical case $\delta n_i \ll 1$.
We calculate $\delta n_i$ by requiring that:
\begin{equation}
\int_0^{\delta \theta_i}\! d\theta\ 
\frac{d\rho}{d\theta}(\theta,\delta n_i)= 0.683\,,
\label{ddd}
\end{equation}
for 
%motivated by the fact that 
the 1 sigma error $\delta \theta_i$ corresponds 
to the $\approx 0.683$ confidence level.
For small $\delta n_i$, we get easily:
\begin{equation}
\delta n_i
\simeq 0.660\ \delta\theta_i \times  (1-\delta\theta^2_i/24).
\label{er1}
\end{equation}
Typically the first term provides
an adequate approximation for the quantity
we search, $\delta n_i$.

Now, we discuss a possible improvement of the 
Gaussian assumption for the distribution $G_2$.
Experimental investigations of
the Super-Kamiokande collaboration~\cite{nim} have shown that
the tails of the angular distribution 
fall slower than $\exp(- \mbox{cte}\cdot
\theta^2)$  and resemble more closely
$\exp(- \mbox{cte}\cdot \theta)$; compare also with App.~C
of~\cite{raff}.
This suggests to release the Gaussian approximation
for the distribution on the directions  and to replace 
$G_2\to$ $\exp(-|\hat{n}-\hat{n}_i|/\delta n_i)$. Thus, the
distribution over the cosine becomes:
\begin{equation}
\frac{d\rho}{d c}=\frac{1}{N_2\;\delta n^2_i}
\exp\! \left[ \!-\frac{\sqrt{2(1-c)}}{\delta n_i}\right],
\label{aaa2}
\end{equation}
where the proportionality constant
\begin{equation}
N_2=1-(1+2/\delta n_i)\exp(-2/\delta n_i),
\end{equation}
is again close to one for small $\delta n_i$.
By imposing the condition (\ref{ddd}) and
considering again the limit of small $\delta n_i$
we calculate the new expression for $\delta n_i$, obtaining:
\begin{equation}
\delta n_i  \simeq 0.424\  \delta\theta_i \times
(1-\delta\theta^2_i/24). \label{er2}
\end{equation}
where, as in Eq.~(\ref{er1}),
the first term is typically sufficient.

\begin{figure}[t]
\centerline{
\includegraphics[width=0.4\textwidth,angle=0]{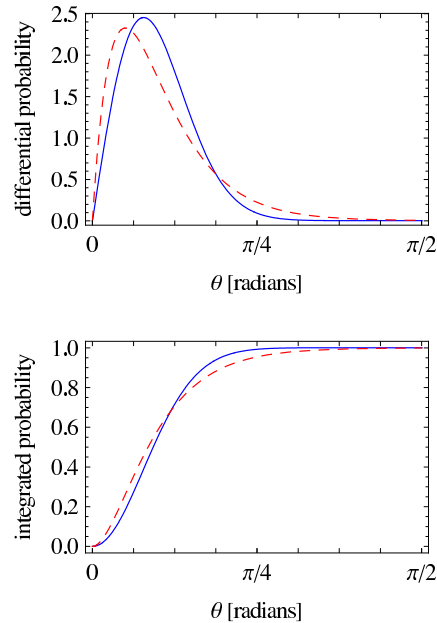}
}
\caption{
Angular distribution (top) and 
related cumulative distribution (bottom) as a function of 
the angle $\theta$. 
The continuous curves show Eq.~(\ref{aaa1})
and the dashed curves show Eq.~(\ref{aaa2}).
We here assumed $\delta \theta=83^\circ \sqrt{\mbox{MeV}/E}$, 
as appropriate for Super-Kamiokande \cite{nim}
and a typical energy of an
elastic scattering event, $E=15$ MeV, so that $\delta \theta =0.374$.
\label{fig1}}
\end{figure}

The two distributions 
are depicted in Fig.~\ref{fig1} for a specific value
of $\delta\theta_i$.
It is worthwhile to note various 
features of Eqs.~(\ref{aaa1}) and (\ref{aaa2}):
\begin{enumerate}
\item When considered as functions of the direction
we see that they both depend only $|\hat{n}-\hat{n}_i|$ and have a 
maximum for $\hat{n}=\hat{n}_i$, as it should be.
\item It is easy  to  treat them analytically, which is a welcome property to
use them in a likelihood.
\item  It is simple to study their limit  for small $\delta n_i$ by replacing
$\sin\theta\to \theta$ and $\cos\theta\to
1-\theta^2/2$, which makes their analytical
treatment even simpler.\footnote{When
only the second replacement is done, Eq.~(\ref{aaa1}) coincides
with the form commonly used in the literature \cite{danuta}
and Eq.~(\ref{aaa2}) practically coincides with the form given in \cite{raff}.
In fact, the exponential term in
Eq.~C1 of \cite{raff} can be neglected in comparison
to the linear term $x$ for all relevant energies.}
\item For a fixed $\delta\theta_i$,
we see that $\delta n_i$ is smaller in the second case;
thus, the  maximum at $\cos\theta=0$ is higher in the second case.
\item The most probable angle is $\theta\simeq \delta n_i$ 
in both cases; thus it is smaller in the second case.

\end{enumerate}

A choice between Eq.~(\ref{aaa1}) and Eq.~(\ref{aaa2}) 
(or other reasonable approximations)
is not critical for the analysis of SN1987A in view
of the limited event sample. However, the use of an appropriate
distribution is potentially important for the analysis of the elastic
scattering events from a future supernova
in a water \v{C}erenkov detector.

\subsection{From the idealized to the actual likelihood}
We are now in the position to provide a concrete and useful
expression for the likelihood that takes into account 
the reported information on the data and on the
detector response.

We recall that:\\
1) The signal is expected to be uniformly distributed inside the detector. 
The angular dependence of the signal arises from the angular distribution 
of positrons produced by $\bar\nu_e p\to n e^+$. This can be expressed as a function of
the angle $\hat{n}\hat{n}_*$ between the direction $\hat{n}$ of the produced 
positron and the direction $\hat{n}_*$ of the SN1987A.
We can thus replace in  
Eq.~(\ref{alfin}):
\begin{equation}
S(t, \mathbf{x} ) \to \frac{S(t,E,\hat{n}\hat{n}_* )}{2\pi V}
\end{equation}
where 
$V$ is the volume of the detector, the product $\hat{n}\hat{n}_*$
can be expressed
through Eqs.(~\ref{est}) and (\ref{nst}) as:
\begin{equation}
\hat{n}\cdot \hat{n}_*= \cos \theta_i \cos\theta + \sin \theta_i
\sin\theta \cos\varphi, \label{glia}
\end{equation}
and the factor $2\pi$ accounts for the fact that positron directions 
are uniformly distributed with respect to rotations around $\hat{n}_*$.\\
2) We assume that the background does not depend on time, direction and position. 
We indicate the total background counting rate as a function of the energy with  $\bar{B}(E)$ and we
replace in Eq.~(\ref{alfin}):
\begin{equation}
B(\mathbf{x_i}) \to \frac{\bar{B}(E)}{4\pi V}\; ;
\end{equation}
3) The average efficiency of the detector as a function of 
the energy $\bar{\eta}(E)$ is known. We assume that the efficiency does 
not depend on time, position and direction, so that we can replace in 
Eq.~(\ref{alfin}):
\begin{equation}
\eta(E,\hat{n},\vec{r})\to\bar\eta({E})\; ;
\end{equation}
4) The errors on the energy $\delta E_i$ 
and on the direction $\delta \theta_i$ in the neighbourhood 
of the given datum are known. We additionally indicate by $\delta r_i$ the 
value of the error on the position, on which we have 
only limited information, $\delta r_i\sim 1$ m at 10 MeV.

At this point, we have all the elements to write the concrete
form of the likelihood. Integrating away the Gaussian
on the positions $G_3$ and omitting the constant factor
$1/(2\pi V)^{N_{\rm ev}}$ we get from Eq.~(\ref{alfin}):
\begin{equation}
\begin{array}{ll}
\mathcal{L}=&e^{-\int_T\! dt ~dE ~dc\ \bar{\eta}(E) S(t,E,c)}
\prod_{i=1}^{N_{\rm ev}} \Big[\ \frac{\bar{B}(E_i)}{2}\ +  \\
 &\int  \bar\eta(E)  ~dE ~G_1(E-E_i,\delta E_i)
\!
\int\! \frac{d\varphi}{2\pi} \int\!\! dc \frac{d\rho}{dc}(c,\delta n_i)
\times\\
&  S(t_i, E,\ c_i\ c  +s_i\; s\; c_\varphi)\ \Big],
\label{prat}
\end{array}
\end{equation}
where Eq.~(\ref{glia}) has been rewritten using the intuitive
shorthands $\cos\theta\to c$, $\sin\theta_i\to s_i$,  etc.

\section{Comparison with the literature\label{four}}

Here, we compare our likelihood, Eq.~(\ref{prat}),
with certain other likelihoods present in the recent
literature and currently used for
the analysis of SN1987A events.

\subsection{Jegerlehner, Neubig and Raffelt~\cite{neubig}}
The first likelihood is Eq.~(15) of~\cite{neubig}:
\begin{equation}
\mathcal{L}^{JNR} =C e^{-\int_0^\infty n(E) dE}
\prod_{i=1}^{N_{obs}} n(E_i).
\end{equation}
This is an approximation of our likelihood, in that
the background has been neglected and the time and angular
distribution are integrated (averaged) over; in other words, only
the energy distribution is considered. However, this expression is
in direct correspondence with Eq.~(\ref{prat}), when $B\to 0$ and
$\delta \theta_i\to 0$, and if we take as the definition of
$n(E_i)$ the one given in Eqs.~(18,19,21) of \cite{neubig}.
Furthermore, the expression of~\cite{neubig} agrees with
Eq.~(\ref{l1}).

\subsection{Lamb and Loredo \cite{ll}}

The other likelihood that we consider is the one
advocated by Lamb and Loredo \cite{ll}. 
This is given by their Eq.~(3.18)
which, rewritten in our notations, reads
\begin{equation}
\begin{array}{l}
\mathcal{L}^{LL}= e^{-\int_T dt dE
\bar{\eta}(E) S(E,t)} \times \\
\ \ \ \ \ \ \ \ \ \ \ \   \prod_{i=1}^{N_{\rm ev}} \left[ \int dE
\mathcal{L}_i(E) S(E,t_i) + \bar{B}(E_i) \right],
\end{array}
\label{LLL}
\end{equation}
where we neglected dead-time, as appropriate for Kamiokan\-de-II,
and dropped the information about the angular
distribution.
Quoting \cite{ll}:
{\em Our derivation of the likelihood function reveals
errors in previous attempts to account for the energy dependence
of the efficiencies of the neutrino detectors; we show that these
errors significantly corrupt previous inferences.}
Indeed, the likelihood advocated by LL
has been shown to have an important impact for the analysis of data
also by similar and independent analyses of SN1987A observations \cite{us}.

\medskip
We would like, however, to draw the discussion on the correctness
of the likelihood of LL. We see that the LL expression,
Eq.~(\ref{LLL}), coincides with our Eq.~(\ref{prat}) {\em only if
we identify} the function $\mathcal{L}_i$ with the energy
response function of the detector:
\begin{equation}
\mathcal{L}_i(E)\stackrel{(?)}{=} G_1(E-E_i ,\sigma_i)
\bar\eta(E), \label{right}
\end{equation}
where $G_1$ is the Gaussian smearing of Eq.~(\ref{cesent}).

This is not the case for LL who instead claim (see their Eq.~(3.21)):
\begin{equation}
\mathcal{L}_i(E)\stackrel{(!)}{=} G_1(E-E_i ,\sigma_i)
\Theta(E-E_0), \label{wrong}
\end{equation}
where $E_0$ is assumed to be the maximum energy
where the efficiency vanishes (i.e., the minimum detectable
energy) and $\Theta$ is the step function.

The only special case in which
the LL likelihood coincides with our result is when the
average efficiency is assumed
to be a step function $\bar\eta(E)=\Theta(E-E_0)$.
In general, this is not the case and
the efficiency is a continuously growing function of the energy.
The LL likelihood therefore incorrectly biases the analysis in favor 
of low energy events. The quantitative effect of this bias on data analysis 
will be discussed further  in Sect.~\ref{qqq}. 

\medskip
The above remarks amount to the consideration that the 
likelihood of Lamb and Loredo does {\em not} follow 
from the formal construction described in Sects.~\ref{prim} and \ref{thir}.
However, it is instructive to point 
out more directly the profound 
principle problem of the LL likelihood.

We begin noting that Eq.~(\ref{LLL}) has been derived 
by omitting constant terms from
\begin{equation}
\begin{array}{l}
\mathcal{P}^{LL}= e^{-\int_T dt dE
[\bar{\eta}(E) S(E,t) + \bar{B}(E)]} \times \\
\ \ \  \prod_{i=1}^{N_{\rm ev}} \left[ \int dE
\mathcal{L}_i(E) S(E,t_i) + \bar{B}(E_i) \right]dt_idE_i,
\end{array}
\label{PLL}
\end{equation}
which should represents the {\em probability} that a given
experimental result is obtained. 
This expression is supposed to
have a general validity. Then consider a simple 
limiting case: only one bin, with dimension
$\Delta t \times \Delta E$ and with energy
above $E_0$; no background, $B(E)\equiv 0$; a
constant signal, $S(E,t)\equiv S$;
a constant efficiency, $\overline{\eta}(E)\equiv \eta$;
a perfect energy resolution, $\delta E_i\rightarrow 0$;
a very small expected number of events,
$n \equiv \eta S \Delta t \Delta E \ll 1$.
In these assumptions, the most probable outcome is 
the case $N_{\rm ev}=0$, followed by the case $N_{\rm ev}=1$; 
the probability of other possible results is negligible.
{}From Eq.~(\ref{PLL}), we calculate the probabilites 
of the cases when no event and one event are observed: 
$\mathcal{P}^{LL}_0=1-n$ and $\mathcal{P}^{LL}_1=n/\eta$ 
respectively. Their sum 
violates the basic principle according to which the sum of the
probability for all possible results should be equal to one.

\subsection{Loredo \cite{loredo} and  Bernstein et al.\ \cite{bernstein}}

We comment here on a likelihood that was {\em not} proposed
for the analysis of supernova neutrinos, but that it
is strictly connected with the previous one.

The same position as in Eq.~(\ref{wrong}) was made
in \cite{loredo}, where Loredo defines the quantity
$\ell_i(m)=p(d_i|m)$.
This quantity, that evidently corresponds to the quantity 
$\mathcal{L}_i$ discussed above, is claimed to be independent on
the detection efficiency.
This position caused the criticism of
Loredo \cite{loredo} to the likelihood
advocated for the analysis of trans-neptunian objects
of Bernstein {\em et al.}\ 2004 \cite{bernstein}
(see Eq.~(A8) of~\cite{bernstein}).
Again, we find the position of \cite{loredo} unjustified,
while we agree with \cite{bernstein}. 
In particular, Eq.~(A4) of \cite{bernstein} 
expresses the statement that the response 
function {\em does} contain the detection efficiency;
it corresponds strictly to our Eq.~(\ref{edef}).

\subsection{Pagliaroli et al.\ \cite{us}}
Finally, in the analyses performed by some of us \cite{us},
Eq.~(\ref{prat}) was used, simplified to the case
$\delta n_i\to 0$ to take into account the mild
angular dependence of the $\bar\nu_e p\to n e^+$ reaction.
This simplification does not affect significantly the
analysis of SN1987A events.

\subsection{Numerical comparison of the likelihoods \label{qqq}}
The bias implied by the likelihood of ref.~\cite{ll}
is numerically important already for the analysis of SN1987A events,
as found in \cite{ll} and confirmed by \cite{us}.  

\begin{figure}[t]
\centerline{
\includegraphics[width=0.34\textwidth,angle=270]{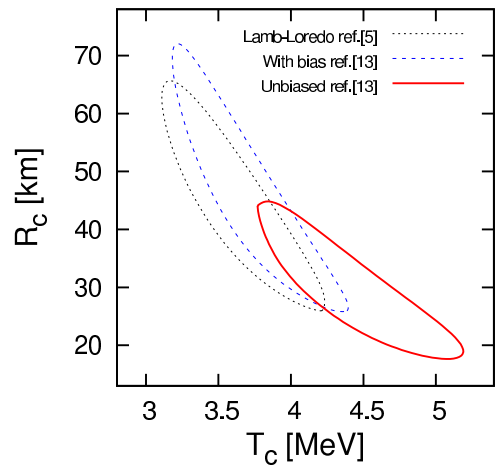}}
\caption{Comparison of the joined 68\% C.L.~(2 dof) regions for the cooling radius $R_c$ and the
initial temperature $T_c$ obtained from two analyses of SN1987A data based on the exponential cooling model.
When Eq.~(\ref{wrong}) is adopted, there is a good agreement 
between the results of \cite{ll} (dotted line) and those of \cite{us} (dashed line); the small 
discrepancies can be ascribed to  
the different statistical procedures (Bayesian in \cite{ll} and frequentist  in \cite{us}) and 
to a different numerical treatment of the data.
The effect of switching from Eq.~(\ref{wrong}) (dashed line) to 
Eq.~(\ref{right}) (continuous line) is much more significant \cite{us}.  
\label{fig2}}
\end{figure}

In order to illustrate this point better, we recall 
certain results obtained in the previous analyses. 
Let us begin by considering the conventional exponential 
cooling model, in which the $\bar\nu_e$ temperature
decreases exponentially with the time and the neutrino-radius $R_c$ is
constant. As evident from Fig.~\ref{fig2}, the use of Eq.~(\ref{wrong}) rather than
Eq.~(\ref{right}) leads to important differences on the inferred values of the
parameters. We note in particular that the (well-known) difference between
$R_c$ and the expected size of the neutron star radius, $R_{\mbox{\tiny
ns}}\sim 15$~km, is amplified when we adopt 
Eq.~(\ref{wrong}) (i.e., when we bias the analysis). This outcome 
can be easily understood: the bias in favor of low energy events implies
that $T_c$ (that is proportional to the average energy of the electron antineutrinos) will 
decrease; thus, $R_c$ has to increase to keep the number of events constant.

\medskip
The bias will be even more important for the
analysis of a future galactic supernova, since the number 
of collected events will be much larger and the errors on the parameters are expected to
scale as the square root of the number of the events.   

Indeed, the analysis of the 29 events collected by Kamiokande-II, IMB and Baksan
in an extension of  the exponential cooling model leads to 
$R_c=16^{+9}_{-5}$ km and $T_c=4.6^{+0.7}_{-0.6}$ MeV  \cite{us}.
A recent analysis of simulated events from a future supernova, 
that assumed the same antineutrino emission model 
(that includes an initial phase of intense emission),
the same central values as found from SN1987A
(in particular, $R_c=16$ km and $T_c=4.6$ MeV), 
and a supernova located at a distance of 20 kpc (i.e., a data set 
30 times larger), leads 
to the conclusions that the parameters are correctly 
reconstructed when we use Eq.~(\ref{right}). 
Moreover, when we combine the results of the simulations, we can estimate the 
average values of the parameters and of their expected errors: 
$R_c=15.4 \pm 0.9 $ km and $T_c=4.6\pm 0.1$ MeV \cite{pg}. 

The comparison with the values from SN1987A 
reveals that the errors are expected to decrease by about six times, 
which is similar to the improvement
that we can ascribe to the increased number of data.
We are lead to the conclusion that, after a future galactic supernova, 
the allowed regions in Fig.~\ref{fig2} should shrink 
by a similar factor in linear scale, making the effect of the bias 
much more important.

%% Elastic scattering is discussed in \cite{raff}, \cite{prd,abc}. 
%% In \cite{abc}, the triple integral of the
%% elastic scattering interaction rate 
%% in Eq.~(\ref{prat}) has been 
%% approximated using: 
%% \begin{equation}
%% \begin{array}{l}
%% \int d\hat{n} 
%% S(t,E_e,\hat{n}\hat{n}_*) G_2(\hat{n}-\hat{n}_i,\delta n)\approx \\[1ex]
%% 2\pi S(t,E_e) G_2(\hat{n}_*-\hat{n}_i,\delta n)
%% \end{array}
%% \end{equation}
%% where for any type of 
%% neutrino, $\nu=\nu_e, \bar{\nu}_e, \nu_\mu, \bar{\nu}_\mu, 
%% \nu_\tau, \bar{\nu}_\tau$, 
%% the differential rate of interactions is given by:
%% \begin{equation}
%% S(t,E_e)=\int_{E_{\mbox{\tiny min}}}^\infty 
%% \!\!\!\!\!\!\! dE\ N_{e}\ 
%% \frac{d\Phi_\nu}{dtdE}(t,E)\ \frac{d\sigma_\nu}{dE_e}(E,E_e).
%% \end{equation}
%% $N_e$ is the number of target electrons, 
%% $d\Phi_\nu/dtdE$ the differential flux of neutrinos of type $\nu$,
%%  $d\sigma_\nu/dE_e$ the elastic scattering cross section and 
%% $E_{\mbox{\tiny min}}(E_e)=(p_e+ E_e-m_e)/2$.

\section{Summary}
We constructed the general likelihood
for supernova data analysis, Eq.~(\ref{alfin}), 
and specified it to the analysis of SN1987A, Eq.~(\ref{prat}).
We have compared this likelihood with other forms advocated
in the scientific literature. 
While our likelihood
is a generalization of the likelihoods traditionally adopted
for the analysis of SN1987A events (or in general for the study
of rare processes), it is in disagreement with other ones.
Reasons and consequences
of these disagreements are discussed.

\section*{Acknowledgments}
We thank A.~Dighe, E.~Lisi, D.~Montanino
and F.~Terranova for useful discussions. 
This work was partly supported by 
High Energy Astrophysics Studies contract number ASI-INAF I/088/06/0;
MIUR grant for the Projects of National Interest 
PRIN 2006 ``Astroparticle Physics''; European 
FP6 Network ``UniverseNet'' MRTN-CT-2006-035863. 
F.R.T.\ thanks CAPES (grant number 3247-08-2) 
for financial support and INFN Gran Sasso for hospitality.

\end{document}